\documentclass{sig-alternate-2013}
\newfont{\mycrnotice}{ptmr8t at 7pt}
\newfont{\myconfname}{ptmri8t at 7pt}
\let\crnotice\mycrnotice%
\let\confname\myconfname%

\conferenceinfo{RecSysChallenge'14,}{October 10, 2014, Foster City, CA, USA.}

\usepackage{amssymb}
\usepackage{graphicx}
\usepackage{array}
\usepackage{multirow}
\usepackage{rotating}

\newcolumntype{C}[1]{>{\centering\let\newline\\\arraybackslash\hspace{0pt}}m{#1}}
\newcolumntype{L}[1]{>{\raggedright\let\newline\\\arraybackslash\hspace{0pt}}m{#1}}

\begin{document}
%

\title{Regression and Learning to Rank Aggregation\\for User Engagement Evaluation}
%
%
%
%
%
\numberofauthors{1} 
\author{\alignauthor Hamed Zamani, Azadeh Shakery, and Pooya Moradi \\
\affaddr{School of Electrical and Computer Engineering, College of Engineering,} \\ \affaddr{University of Tehran, Tehran, Iran} \\
\email{\{h.zamani, shakery, po.moradi\}@ut.ac.ir}
}
\date{28 July 2014}

\maketitle
\begin{abstract}
User engagement refers to the amount of interaction an instance (e.g., tweet, news, and forum post) achieves. Ranking the items in social media websites based on the amount of user participation in them, can be used in different applications, such as recommender systems. In this paper, we consider a tweet containing a rating for a movie as an instance and focus on ranking the instances of each user based on their engagement, i.e., the total number of retweets and favorites it will gain.

For this task, we define several features which can be extracted from the meta-data of each tweet. The features are partitioned into three categories: user-based, movie-based, and tweet-based. We show that in order to obtain good results, features from all categories should be considered. We exploit regression and learning to rank methods to rank the tweets and propose to aggregate the results of regression and learning to rank methods to achieve better performance.

We have run our experiments on an extended version of MovieTweeting dataset provided by ACM RecSys Challenge 2014. The results show that learning to rank approach outperforms most of the regression models and the combination can improve the performance significantly.
\end{abstract}




\section{Introduction}
Twitter is an online social information network which has become tremendously popular in the past few years \cite{Kywe:2012}. Millions of users are sharing rich information using social media sites, such as Twitter, which can be used by social recommender systems \cite{Guy:2011}. Item providers often let users express their opinion about an item in social networks. For instance, users can give a rating to each movie in Internet Movie Database (IMDb) website\footnote{http://imdb.com} and also share it in Twitter. This intensifies the importance of considering social media sites for recommendation and information filtering systems \cite{Uysal:2011}.

Product rating prediction is a traditional recommender system problem which has been studied extensively in the literature \cite{Ekstrand:2011,Nguyen:2009,Oghina:2012}. One important issue in recommender systems is the engagement which can be gained by the users' comments/opinions. When users share their comments on different items, the amount of user interactions achieved by each comment can be used to improve the quality of recommender systems. In this paper, we focus on ranking these comments by their engagements.

We focus on movie ratings tweeted by IMDb users in Twitter. Hereafter, we use the word ``engagement" as the user interaction which is expressed by adding up the number of retweets and favorites a tweet has gained. Our purpose is to rank the tweets of each user, each containing a rating for a movie in IMDb, by their engagements.

For this task, we first extract several features from the tweets. The features are categorized into three groups: user-based, movie-based, and tweet-based. It should be noted that the content of the tweets are hidden and there is no textual feature among our defined features. Then, we propose two different supervised approaches in order to rank the tweets. The first approach tires to predict the tweets engagements globally. In other words, although our purpose is to sort the tweets of each user, we consider tweets of all the users together and then try to predict the tweets engagements. We can then extract the sorted list of each user from the global ranked list. Therefore, we fit regression models to predict the engagement of each tweet. In the second approach, for each user, we rank the tweets by their engagement without predicting the engagements. To this aim, we use learning to rank approach which is extensively exploited in information retrieval, natural language processing, and recommender systems. Learning to rank methods rank the tweets for each user. In contrary to regression models which try to predict the engagements by considering all the tweets together, learning to rank methods emphasize on maximizing an objective function for each user. According to the different points of view of regression and learning to rank methods, we further propose to aggregate the results obtained by different regression and learning to rank methods to improve the performance.

In the experiments, we use an extended version of MovieTweetings dataset \cite{Dooms:2013} provided by ACM RecSys Challenge 2014 and report the results of a number of state-of-the-art regression and learning to rank methods, separately. We further discuss the aggregation of the results of these two approaches. The experimental results show that although the results of regression methods are not so impressive, aggregation of regression and learning to rank methods improves the results significantly.


\section{Related Work}
\label{sec:relwork}
The problem of engagement prediction or online participation has been studied from different points of view in news websites, social networks, and discussion forums. Several machine learning algorithms have been used in the literature for this task.

To address the problem of engagement prediction, several features have been proposed for training a model. Suh et al. \cite{Suh:2010} have provided an analysis on the factors impacting the  number of retweets. They have concluded that hashtags, number of followers, number of followees, and the account age play important roles in increasing the probability of the tweets to be retweeted. Zaman et al. \cite{Zaman:2010} have trained a probabilistic collaborative filtering model to predict the future retweets using the history of the previous ones.

Linear models have been used in some other studies to predict the popularity of videos on YouTube by observing their popularity after regular periods \cite{Szabo:2010}. Petrovic et al. \cite{Petrovic:2011} have proposed a passive-aggressive algorithm to predict whether a tweet will be retweeted or not.

Recognizing popular messages is also one of the similar problems which is used for breaking news detection and personalized tweet/content recommendation. Hong et al. \cite{Hong:2011} have formulated this task as a classification problem by exploiting content-based features, temporal information, meta-data of messages, and the users social graph.

Predicting the extent to which a news is going to be breaking or how many comments a news is going to gain is one of the engagement prediction problems. Tatar et al. \cite{Tatar:2012} have analyzed a news dataset to address this problem. They have focused on sorting the articles based on their future popularity and they have proposed to use linear regression for this task.

It is worth noting that ranking instances is one of the problems which has been extensively studied in information retrieval, natural language processing, and machine learning fields \cite{Li:2011}. To solve a similar problem, Uysal and Croft \cite{Uysal:2011} have proposed ``Coordinate Ascent learning to rank" algorithm to rank tweets for a user in a way that tweets which are more likely to be retweeted come on top. They have also worked on ranking users for a tweet in a way that the higher the rank, the more likely the given tweet will be retweeted. Several learning to rank algorithms have been proposed in the literature. Moreover, there are some supervised and unsupervised ensemble methods to aggregate different rankings, such as Borda Count \cite{Aslam:2001} and Cranking \cite{Lebanon:2002}. Previous studies show that in many cases, ranking aggregation methods outperform single ranking methods \cite{Chapelle:2011,Li:2011}.

\vfill
\pagebreak

\section{Methodology}
\label{sec:method}
In general, our idea is to extract a number of features for each tweet and then try to learn machine learning based models on the training data. Then, for each user in test data, we apply the learned model to rank his/her tweets based on their engagements. In this section, we first introduce the features, and then we propose some machine learning approaches to rank the tweets based on their engagements. We also try to aggregate the results of these different techniques to improve the performance. In the following subsections, we explain our methodology in details.

\subsection{Features}
\label{sec:features}
Each tweet contains the opinion of a user about a specific movie. We partition the features extracted from each tweet into three different categories: user-based, movie-based, and tweet-based features. Overall, we extract several features from each tweet T tweeted by user U about movie M. User-based features give us some information about the user who has tweeted his/her opinion about a specific movie. These features are not tweet-specific and they are equal for all tweets of each user. The total number of followers of U is an example of user-based features. Movie-based features only include information about movie M, e.g., the total number of tweets about movie M. Tweet-based features contain specific information of tweet T. This information may also contain the opinion of user U about movie M. The time and language of a tweet are two examples of tweet-based features.

The name and description of the extracted features are shown in Table \ref{tab:features}. These features are extracted for each tweet T. We specify the category of the features and also their type; ``N", ``C", and ``B" are used for numerical, categorical, and boolean types, respectively. It should be noted that the feature values are normalized using z-score normalization method.

We also perform feature selection to improve the performance and also to analyse the effectiveness of the proposed features. We exploit backward elimination for feature selection. The bolded features in Table \ref{tab:features} are those that are retained after performing feature selection. We discuss the selected features in Subsection \ref{sec:res}

\bgroup
\def\arraystretch{1.5}
\begin{table*}
\centering
\caption{Extracted features from each tweet T tweeted by user U about movie M}
\label{tab:features}
\begin{tabular}{|p{0.75cm}|p{3cm}|C{0.85cm}|p{11cm}|} \hline
\multicolumn{1}{|c|}{\bfseries Cat.} & \multicolumn{1}{c|}{\bfseries Feature Name} & \textbf{Type} & \multicolumn{1}{c|}{\bfseries Description}\\ \hline
\multirow{18}{*}{\begin{sideways}User-based\end{sideways}} & \textbf{Number of followers} & N & The total number of users who are following user U in Twitter. \\ \cline{2-4}
 & Number of followees & N & The total number of users who are followed by user U in Twitter.\\ \cline{2-4}
 & \textbf{Number of tweets} & N & The total number of tweets written by user U.\\ \cline{2-4}
 & \textbf{Number of IMDb tweets} & N & The total number of tweets tweeted by user U using IMBD about different movies.\\ \cline{2-4}
 & Average of ratings & N & The average of ratings provided by user U about different movies in IMDb.\\ \cline{2-4}
 & Number of liked tweets & N & The total number of tweets which are liked by user U.\\ \cline{2-4}
 & Number of lists & N &The total number of Twitter lists which user U is involved in.\\ \cline{2-4}
 & Tweeting frequency & N & The frequency of tweets written by user U in each day.\\ \cline{2-4}
 & Attracting followers frequency & N & The frequency of attracting followers per day. This feature is calculated by dividing the total number of followers by the membership age of user U in Twitter in terms of number of days.\\ \cline{2-4}
 & Following frequency & N &The frequency of following different users by user U per day.\\ \cline{2-4}
 & Like frequency & N &The frequency of liking tweets by user U per day.\\ \cline{2-4}
 & Followers/Followees & N & The total number of followers of user U divided by the total number of his/her followees.\\ \cline{2-4}
 & \textbf{Followers-Followees} & N & The difference between the total number of followers and followees of user U.\\ \hline
\multirow{4}{*}{\begin{sideways}Movie-based\end{sideways}} & Number of tweets about M & N & The total number of tweets tweeted using IMDb about movie M. This feature shows how much movie M is rated by different users around the world in IMDb. \\ \cline{2-4}
 & \textbf{Average rating of M} & N & The average of ratings reported by different users for movie M.\\ \hline
\multirow{16}{*}{\begin{sideways}Tweet-based\end{sideways}} & \textbf{Rate} & N & The rating provided by user U for movie M. This rating is a positive integer up to 10. \\ \cline{2-4}
 & \textbf{Mention count} & N & The total number of people who are mentioned in tweet T.\\ \cline{2-4}
 & Number of hash-tags & N &The total number of hash-tags used in tweet T.\\ \cline{2-4}
 & Tweet age & N & The age of tweet T in terms of number of days.\\ \cline{2-4}
 & Membership age until now & N & The number of days from when user U registered in Twitter until when tweet T is tweeted.\\ \cline{2-4}
 & \textbf{opinion difference} & N & The difference between the rate tweeted by user U for movie M and the average of rates given by different users about movie M.\\ \cline{2-4}
 & Hour of tweet & C & The hour when tweet T is tweeted. This feature is an integer between 0 and 23.\\ \cline{2-4}
 & Day of tweet & C & The day of week which tweet T is tweeted.\\ \cline{2-4}
 & Time of tweet & C & The part of the day that tweet T is tweeted. We have partitioned each day into four parts.\\ \cline{2-4}
 & \textbf{Holidays or not} & B & This feature give us whether tweet T is tweeted on holidays or not.\\ \cline{2-4}
 & \textbf{Same language or not }& B & This feature illustrates whether tweet T is tweeted in the same language as the default language of user U or not.\\ \cline{2-4}
 & \textbf{English or not} & B & This feature tells us whether tweet T is tweeted in English or not.\\ \hline

\end{tabular}
\end{table*}
\egroup

\subsection{Machine Learning Techniques for User Engagement Ranking}


In this subsection, we propose two different learning based approaches to rank the tweets of each user based on their engagements. The first approach is predicting the engagement of tweets, globally. In other words, for predicting the engagement of tweets of a user, we consider the tweets of all users for training the model and not only the tweets of the user. To this aim, we use regression models to predict the engagement of each tweet. The next approach is to rank the tweets for each user without predicting their engagements. We exploit learning to rank methods to rank the tweets of each user, which focus on ranking the tweets of each user individually and try to maximize a given objective function for each user. Finally, we propose a supervised method to aggregate the regression and learning to rank results using supervised Kemeny approach \cite{Agarwal:2012}. In the following, we explain our proposed methods in details.

%

\vfill
\pagebreak

\subsubsection{Regression}
To rank the tweets of each user based on their possible engagements, we can first predict the engagement of each tweet and then sort the tweets by their predicted values. To predict the engagements, we propose to train regression models by using the features defined in Subsection \ref{sec:features} as the features and the engagements as the labels. Then, we apply the learned model on the same extracted features from the test set.

To create the regression model, we exploit Extremely Randomized Trees (also known as Extra-Trees) \cite{Geurts:2006}, Bayesian Ridge Regression \cite{Mackay:1992}, and Stochastic Gradient Descent Regression (SGDR) \cite{Leon:2004}. Extra-Trees are tree-based ensemble regression methods which are successfully used in several tasks. In Extra-Trees, when a tree is built, the node splitting step is done randomly by choosing the best split among a random subset of features. The results of all trees are combined by averaging the individual predictions. SGDR is a generalized linear regression model that tries to fit a linear model by minimizing a regularized empirical loss function using gradient descent technique.

\subsubsection{Learning to Rank}
Instead of predicting the exact engagements, we can rank the tweets directly, without predicting the engagements of each tweet. Learning to Rank (LTR) methods are machine learning techniques which try to solve ranking problems \cite{Li:2011}. LTR methods have been widely used in many different areas such as information retrieval, natural language processing, and recommender systems \cite{Karatzoglou:2013,Li:2011}. LTR methods train a ranking model and use the learned model to rank the instances using several features which are extracted from each instance.

To build our LTR model, we consider a number of ranking algorithms which are among state-of-the-art in many test collections: ListNet \cite{Cao:2007}, RankingSVM \cite{Joachims:2002}, AdaRank \cite{Xu:2007}, RankNet \cite{Burges:2005}, LambdaRank \cite{Burges:2007}, and ListMLE \cite{Xia:2008}. ListNet is a probabilistic listwise approach to solve ranking problems, which exploits a parameterized Plackett-Luce model to compute different permutations. Ranking SVM is a pairwise ranking approach which uses SVM classifier in its core computations. The basic idea behind AdaRank is constructing some weak rankers and combining them linearly to achieve a better performance. Although, Ranking SVM creates a ranking model by minimizing the classification error on instance pairs, AdaRank tries to minimize the loss function which is directly defined as an evaluation measure (such as NDCG@10). RankNet is one of the pairwise methods that adopts cross entropy as the loss function. RankNet employs a three layered neural network with a single output node to compare each pairs. LambdaRank is one of the ranking algorithms inspired by RankNet which uses Gradient Descent approach to optimize the evaluation measure. Similar to ListNet, ListMLE is a probabilistic listwise approach to rank instances by maximizing a logarithmic loss function.

\subsubsection{Aggregating Regression and Learning to Rank Outputs}
According to the aforementioned facts, regression and learning to rank techniques take two different points of view into consideration and their results might be totally different. Therefore, by aggregating their results, the performance can potentially be increased.

To aggregate all the mentioned regression and learning to rank results, we use \textit{supervised Kemeny approach} \cite{Agarwal:2012}. Kemeny optimal aggregation \cite{Kemeny:1959} tries to minimize total number of pairwise disagreements between the final ranking and the outputs of all base rankers. In other words, if $r_1,~r_2,~...,~r_n$ represent the outputs of $n$ different rankers, the final ranking $r^*$ is computed as:
\begin{equation*}
r^* = \arg\max_{r}~\{\sum_{i=1}^{n}{k(r,~r_i)}\}
\end{equation*}
where $k(\alpha,~\beta)$ is the Kendall tau distance \cite{Kendall:1938} measured as:
\begin{equation*}
|{(i,~j):~i<j,~\alpha_i>\alpha_j~\wedge~\beta_i<\beta_j}|
\end{equation*}
where $\alpha_i$ denotes the $i^{th}$ position of ranking $\alpha$.

While in Kemeny optimal aggregation all the rankers have the same importance, supervised Kemeny approach assumes that there is a weight for each ranker. In more details, in supervised Kemeny instead of counting the number of disagreements, we use the following equation to compute the final ranking:
\begin{equation*}
r^* = \arg\max_{r}~\{\sum_{i=1}^{n}{k(r,~r_i)}*w_i\}
\end{equation*}
where $w_i$ denotes the weight of $i^{th}$ ranker. To find the weight of each ranker, we propose to perform a Randomized Search \cite{Bergstra:2012}. To this aim, we perform cross validation over training data and find the optimal weight for each ranker.



\section{Experiments}
\label{sec:exp}
In the experiments, we consider an extended version of MovieTweetings dataset \cite{Dooms:2013} which is provided by ACM RecSys Challenge 2014 \cite{Said:2014}.\footnote{http://2014.recsyschallenge.com/} The dataset contains movie ratings which are automatically tweeted by the users of IMDb iOS application. The reported results throughout this work are those obtained on the test set. The evaluation measure is the mean of normalized discounted cumulative gain \cite{Jarvelin:2002} computed for top 10 tweets of each user. We call it \textit{NDCG@10}, hereafter.

In our experiments, we used Scikit-learn library \cite{Pedregosa:2011} for all the regression and feature selection algorithms. To select the parameters of the learning methods, we performed hyper-parameter optimization using Randomized Search \cite{Bergstra:2012} with 5-fold cross validation. For the learning to rank algorithms except AdaRank, we exploited an open source package, named ToyBox-Ranking\footnote{https://github.com/y-tag/cpp-ToyBox-Ranking}. For AdaRank, we used the software developed in Microsoft Research \cite{Xu:2007}.\footnote{http://goo.gl/xycK0h}

\subsection{Experimental Results and Discussion}
\label{sec:res}
In this subsection, we report and discuss the results of different regression and learning to rank methods. We also provide the results obtained by aggregating the regression and learning to rank results using the supervised Kemeny approach.

To show the impact of feature selection, we report the results of regression and learning to rank methods both before and after feature selection. As mentioned before, the bolded features in Table \ref{tab:features} are those retained after performing backward elimination method. The selected features are diffused among all the three feature categories. This shows the importance of using a combination of different kinds of features in this problem. The selected user-based features show how active and popular the user is in Twitter. Interestingly, all the boolean features are selected and none of the categorical features are retained. The reason may be that the values of the boolean features are constant and the difference between them are not a continuous value. So it may be easier and more efficient to use these features. Moreover, for the categorical features, we assign a number to each possible category and the arithmetic difference between these numbers is not informative.

Table \ref{tab:reg} shows the results obtained by different regression algorithms, in terms of NDCG@10. In Table \ref{tab:reg}, ``XT", ``BRR", and ``SGDR" respectively denote Extremely Randomized Trees, Bayesian Ridge Regression, and Stochastic Gradient Descent Regression.

The results reported in Table \ref{tab:reg} demonstrate that feature selection does not help with regression algorithms. In other words, after performing the feature selection, the results of regression models are dropped dramatically. This shows that backward elimination is not sufficient for regression models. According to Table \ref{tab:reg}, there is a considerable difference between the results achieved by different regression models.

\bgroup
\def\arraystretch{1.5}
\begin{table}
\centering
\begin{small}
\caption{Regression results with and without feature selection}
\label{tab:reg}
\begin{tabular}{|C{2cm}|C{2.5cm}|C{2.5cm}|} \cline{2-3}
\multicolumn{1}{c|}{} & \multicolumn{2}{c|}{NDCG@10}\\ \hline
REG method  & REG w/ FS & REG w/o FS\\ \hline
XT & 0.7441384724 & 0.7863435909 \\ \hline
BRR & 0.7541443109 & 0.7759180414 \\ \hline
SGDR & 0.7507494314 & 0.8168741812 \\ \hline
\end{tabular}
\end{small}
\end{table}
\egroup

Table \ref{tab:ltr} shows the results of using several learning to rank methods. The results also include NDCG@10 before and after applying feature selection. The results reported in Table \ref{tab:ltr} emphasize on the importance of using feature selection in learning to rank methods; since after performing feature selection, the results are improved. Therefore, backward elimination method works well for LTR methods. Table \ref{tab:ltr} demonstrates that ListNet performs better than the other LTR methods. Comparing the results of Table \ref{tab:reg} and Table \ref{tab:ltr} shows that all the learning to rank methods outperform all the regression models.

\bgroup
\def\arraystretch{1.5}
\begin{table}
\centering
\begin{small}
\caption{Learning to rank results with and without feature selection}
\label{tab:ltr}
\begin{tabular}{|C{2.1cm}|C{2.5cm}|C{2.5cm}|} \cline{2-3}
\multicolumn{1}{c|}{} & \multicolumn{2}{c|}{NDCG@10}\\ \hline
LTR method  & LTR w/ FS& LTR w/o FS\\ \hline
ListNet & 0.8243394623 & 0.8190048552 \\ \hline
RankingSVM & 0.8225893034 & 0.8169257071 \\ \hline
AdaRank & 0.8182340058 & 0.8153622186 \\ \hline
RankNet & 0.8223464432 & 0.8169752826 \\ \hline
LambdaRank & 0.8209622031 & 0.8126243442 \\ \hline
ListMLE & 0.8217342257 & 0.8174866943 \\ \hline
\end{tabular}
\end{small}
\end{table}
\egroup

Table \ref{tab:agg} represents the results obtained by aggregating the mentioned regression and learning to rank results using supervised Kemeny approach.
To show the importance of considering both regression and learning to rank methods together, we also report the results achieved by aggregating all the LTR methods and all the regression methods, separately. Table \ref{tab:agg} indicates that although most of the results of regression models are far lower than the LTR methods, their aggregation improves the results. It shows that aggregating regression and learning to rank methods achieves better results in comparison with aggregating only LTR methods or regression models. To show that this improvement is significant, we performed 10-fold cross validation over the training data and conducted a statistical significant test (\textit{t-test}) on the improvements of LTRs+REGs over the other methods. The results show that the improvement achieved by LTRs+REGs is statistically significant ($p-value<0.01$).

\bgroup
\def\arraystretch{1.5}
\begin{table}
\centering
\begin{small}
\caption{Ranking aggregation results}
\label{tab:agg}
\begin{tabular}{|C{2.1cm}|C{3cm}|} \cline{2-2}
\multicolumn{1}{c|}{} & \multicolumn{1}{c|}{NDCG@10}\\ \hline
LTRs & 0.8242044953 \\ \hline
REGs & 0.8063031984 \\ \hline
LTRs+REGs & 0.8261454943 \\ \hline
\end{tabular}
\end{small}
\end{table}
\egroup

\section{Conclusions}
\label{sec:conclusion}
In this paper, to rank the tweets of each user based on their engagements, we first defined several features partitioned into three different categories: user-based, movie-based, and tweet-based. We showed that after performing feature selection, the features are selected from all of these categories. Then, we exploited regression and learning to rank methods to rank the tweets of each user by their engagements. Finally, we aggregated the results of all the regression and learning to rank methods using supervised Kemeny approach.

We evaluated our methods on an extended version of MovieTweeting dataset provided by ACM RecSys Challenge 2014. The experimental results demonstrate that feature selection significantly affects the performance. The results also show that however the results of most regression models are far lower than learning to rank methods, their aggregation improves the performance.


\bibliographystyle{abbrv}
\bibliography{sigproc}  

\end{document}